\documentclass[twocolumn,preprintnumbers,amsmath,amssymb,prl,aps]{revtex4}

\usepackage{dcolumn}
\usepackage{bm}

\begin{document}


\title{Exchange interaction parameters and adiabatic spin-wave spectra of
ferromagnets:\\
A ``{\em renormalized} magnetic force theorem''}

\author{P. Bruno}
\email{bruno@mpi-halle.de} \affiliation{Max-Planck-Institut f\"ur
Mikrostrukturphysik, Weinberg 2, D-06120 Halle, Germany}
\homepage{http://www.mpi-halle.de}

\date{24 July, 2002; revised 6 November 2002}

\begin{abstract}
The ``magnetic force theorem'' is frequently used to compute
exchange interaction parameters and adiabatic spin-wave spectra
of ferromagnets. The interest of this approach is that it allows
to obtain these results from a {\em non-self-consistent}
calculation of the {\em (single-electron) band energy} only, which
greatly reduces the computational effort. However, as discussed by
various authors, this approach entail some systematic error. Here,
a ``{\em renormalized} magnetic force theorem'' allowing to remove
this systematic error without increasing significantly the
computational effort is presented. For systems with one atom per
unit cell, it amounts to a simple renormalization of the spin-wave
spectrum. This renormalization greatly improves the agreement
between calculated and experimental Curie temperatures of Fe and
Ni.\\
\\
published in: Phys. Rev. Lett. {\bf 90}, 087205 (2003)
\end{abstract}


\maketitle

The {\em ab initio\/} study of magnetic interactions in magnetic
metals and of interlayer exchange coupling in multilayers has
been the subject of a large number of publications during the
past few years. Many these studies rely explicitly or implicitly
on the use of the ``magnetic force theorem'' (MFT), which allows
to perform the calculations {\em non-self-consistently\/}, and by
taking into account only {\em single-particle energies\/}. This
is of great practical interest for it reduces the computational
effort by several orders of magnitude.

The force theorem was first introduced for the case of
non-magnetic systems \cite{Mackintosh1980}. Extensions to the
case of magnetic systems have been published by Oswald {\em et
al.\/} \cite{Oswald1985}, who focused on the case of magnetic
impurities embedded in a non-magnetic host, and by Liechtenstein
{\em et al.\/} \cite{Liechtenstein1984}, who addressed the case
of exchange interactions and spin-wave spectra of ferromagnetic
systems. However, as mentioned by Liechtenstein {\em et al.} in
their 1984 paper, their MFT yields exact results only in the
limit of infinite magnon wavelength. At finite magnon wavelength,
their prescription entails some systematic error, as emphasized
recently by various authors \cite{Bylander2000, Grotheer2001}.

In the present Letter I present a ``{\em renormalized} MFT'',
which corrects the systematic error entailed by the MFT of
Liechtenstein {\em et al.} \cite{Liechtenstein1984} (hereafter
called the ``{\em bare} MFT''). In the case of periodic systems
with a single site per unit cell, this results in an extremely
simple renormalization of the spin-wave spectrum and Curie
temperature. The Curie temperatures calculated with this
procedure are found to be in excellent agreement with the
experimental ones, both for Fe and Ni.

The problem we want to address here is that of calculating the
energy required to create some static transverse deviation of the
magnetic moments in a ferromagnet. The physical motivation for
this is that, within the adiabatic approximation (i.e., provided
that the characteristic time scale for the dynamics of such
fluctuations is long as compared with the time scale for the
dynamics of band electrons), it provides a good approximation to
the energy of the low-lying magnetic excitations and therefore
allows to address from an {\em ab initio} point of view the
thermodynamics of itinerant ferromagnets. In practice this
results into a mapping of the complicated itinerant ferromagnetic
system onto an effective Heisenberg system having the same
low-lying excitations, but much less degrees of freedom. Let
$n({\bf r})$ and ${\bf m}({\bf r})\equiv m({\bf r}) {\bf u}({\bf
r})$ be, respectively, the charge- and spin-density. For
transition-metal systems, it is usually justified to neglect (as
is almost always done) intra-atomic non-collinearity and
fluctuations (which are generally expected to cost a high
energy). We therefore have ${\bf u}({\bf r}) = {\bf u}_{\bf R}$
if ${\bf r}$ belongs to the atomic cell $\Omega_{\bf R}$ around
atom ${\bf R}$, and we wish to impose a prescribed direction to
the unit vector ${\bf u}_{\bf R}$ of each site ${\bf R}$
\cite{generalization_continuum}.

The proper way of doing this relies on the ``constrained density
functional theory'' of Dederichs {\em et al.}
\cite{Dederichs1984}, which, in the present case, amounts to
introduce some local external field ${\bf B}^\bot_{\bf R}$,
perpendicular to the local magnetic moment axis ${\bf u}_{\bf
R}$, playing the role of Lagrange parameters. As we wish to
constrain two components of the magnetic moment for each site, we
have two Lagrange parameters per site, which are the magnitude and
azimuthal angle of the local constraining field.

In (non-relativistic) spin-density functional theory
\cite{Hohenberg1964}, the ground state energy $E_0$ is obtained
by minimizing the Hohenberg-Kohn functional (HKF)
%
$
{\cal E}_\text{HK}[\rho] = {\cal T}_0[\rho] + {\cal
E}_\text{ext}[\rho] + {\cal E}_H[\rho] + {\cal E}_\text{xc}[\rho]
,
$
%
with respect to the spinor-density $\rho \equiv (n\sigma_0+{\bf
m}\cdot \bm{\sigma})$, where $\sigma_0$ is the unit spinor and
${\bm\sigma}$ the vector-spinor whose components are the Pauli
matrices. The various terms are respectively, the kinetic energy
of a non-interacting system having the same spinor-density, the
potential energy, the Hartree part of the Coulomb energy, and the
exchange-correlation energy. The constrained ground state energy
is found by introducing the new constrained HKF (CHKF)
\cite{Dederichs1984}
%
$
{\cal F}_\text{HK}[\rho, {\bf B}^\bot ]\equiv {\cal
E}_\text{HK}[\rho] + {\cal E}_\text{cons}[\rho,{\bf B}^\bot] ,
$
%
where the constraint term is given by
%
$
{\cal E}_\text{cons}[\rho,{\bf B}^\bot] \equiv - \sum_{\bf R} {\bf
B}^\bot_{\bf R} \cdot \int_{\Omega_{\bf R}} {\rm d}{\bf r} \ {\bf
m}({\bf r}),
$
%
and minimizing ${\cal F}_\text{HK}[\rho, {\bf B}^\bot ]$ with
respect to $\rho({\bf r})$, and to the set ${\bf B}^\bot \equiv \{
{\bf B}^\bot_{\bf R}\}$. As we want to calculate the energy
change associated with some {\em infinitesimal} transverse
fluctuation $\delta{\bf u}_{\bf R}$ away from the ferromagnetic
configuration, we may attempt to use the variational properties
of the CHKF in order to minimize the computational effort. The
difficulty lies in the calculation of the kinetic energy term,
which requires the knowledge of the effective one-electron
spinor-potential $w_\text{eff}[\rho]\equiv
V_\text{eff}\sigma_0-{\bf B}_\text{eff}\cdot\bm{\sigma}$ of
energy eigenvalues $\varepsilon_i(w_\text{eff}[\rho])$ (labeled
in order of increasing energy) and eigenvectors $\psi_i({\bf r})$
which yields $\rho$ as output spinor-density, i.e,
%
$
\rho ({\bf r}) = \sum_{i=1}^{N_\text{el}} \left\{
\left|\psi_i({\bf r})\right|^2 \sigma_0 + \left[ \psi_i^\dag({\bf
r})\bm{\sigma}\psi_i({\bf r})\right] \cdot \bm{\sigma} \right\} .
$
%
The kinetic energy is then given by
%
$
{\cal T}_0[\rho] = \sum_{i=1}^{N_\text{el}}
\varepsilon_i(w_\text{eff}[\rho]) - \int {\rm d}{\bf r}\ \left(
nV_\text{eff}-{\bf m\cdot B_\text{eff}} \right) .
$
%
This {\em implicit} dependence of $w_\text{eff}[\rho]$ upon the
spinor-density $\rho$ is the origin of our difficulties. In fact,
starting from a trial {\em input} potential and magnetic field
and solving the one-electron problem, we are only able to obtain
the value of the HKF for the {\em output} spinor-density, which we
did not know {\em a priori}, but not for some spinor-density
chosen {\em a priori}. This problem can be circumvented by using
an auxiliary energy functional, first introduced by Harris
\cite{Harris1985}. The Harris functional (HF) ${\cal
F}_\text{Harris}[\rho,{\bf B}^\bot]$, in the present context, has
the same form as the CHKF ${\cal F}_\text{HK}[\rho,{\bf B}^\bot]$,
with ${\cal T}_0[\rho]$ replaced by
\begin{equation}
{\cal T}^\prime [\rho,{\bf B}^\bot] \equiv
\sum_{i=1}^{N_\text{el}}
\varepsilon_i(w^\prime_\text{eff}[\rho,{\bf B}^\bot]) - \int {\rm
d}{\bf r}\ \left( nV^\prime_\text{eff}-{\bf m\cdot
B^\prime_\text{eff}} \right) ,
\end{equation}
where the new effective spinor-potential
$w^\prime_\text{eff}[\rho,{\bf B}^\bot]\equiv
V^\prime_\text{eff}\sigma_0-{\bf
B}^\prime_\text{eff}\cdot\bm{\sigma}$ is {\em defined explicitly}
in terms of $\rho$ and ${\bf B}^\bot$ by
\begin{equation}
w^\prime_\text{eff}[\rho,\!{\bf B}^\bot]({\bf r}) \! \equiv \frac{
\delta\! \left( {\cal E}_\text{ext}[\rho] \!\! + \! {\cal
E}_H[\rho] \!\! + \! {\cal E}_\text{xc}[\rho] \!\! + \! {\cal
E}_\text{cons}[\rho,{\bf B}^\bot ]\right)}{\delta\rho({\bf r})} ,
\end{equation}
where the functional derivative is taken at the a priori
prescribed spinor density. This essential difference between
$w^\prime_\text{eff}[\rho,{\bf B}^\bot ]$ and
$w_\text{eff}[\rho]$ allows us to calculate {\em explicitly} the
value of the HF for a spinor-density $\rho$ and constrains ${\bf
B}^\bot$ chosen {\em a priori} (provided we know some suitable
approximation of ${\cal E}_\text{xc}[\rho]$), which constitutes a
great advantage with respect to the CHKF. In view of the
Kohn-Sham theorem \cite{Hohenberg1964}, the HF and CHKF obviously
take the same value $E_0$ for the density $\rho^\star$ and
constrains ${\bf B}^{\bot\star}$ corresponding to the constrained
ground state (solution of the constrained Kohn-Sham equation). In
addition, the HF can be shown to be {\em stationary} (but not
necessarily {\em minimal}, in contrast to the CHKF) with respect
to $\rho$ and ${\bf B}^\bot$ in the vicinity of the constrained
ground state $(\rho^\star ,{\bf B}^{\bot\star})$. The properties
of the HF have been studied by various authors who found that, in
fact, it often yields a better approximation of the ground state
energy, in the vicinity of $(\rho^\star ,{\bf B}^{\bot\star})$,
than the corresponding HKF \cite{Read1989}.

Starting from the ferromagnetic state (with ${\bf u}^0_{\bf
R}={\bf u}^0$ for all sites ${\bf R}$), for which we assume the
self-consistent density $\rho_0 \equiv n_0\sigma_0+ m_0 {\bf u}_0
\cdot \bm{\sigma}$ to be known, we perform some infinitesimal
rotations: ${\bf u}_{\bf R}= {\bf u}^0 + \delta {\bf u}_{\bf R}$.
For the ferromagnetic state, the constraining field vanishes
everywhere. For the rotated state, we approximate the energy by
using the HF ${\cal F}_\text{Harris}[\rho,{\bf B}^\bot]$,
evaluated for a trial input density equal to $\rho_\text{in}
\equiv n_0\sigma_0+ m_0({\bf u_0}+\delta{\bf u})\cdot
\bm{\sigma}$, and for some trial input constrain ${\bf
B}^\bot_\text{in}$ (to be specified later).

If we use the local density approximation (LDA) for the exchange
correlation term (as is almost always done), the only term in our
trial evaluation of the HF which varies with $\delta{\bf u}
\equiv \{\delta {\bf u}_{\bf R}\}$ is the band energy (sum of
eigenvalues $\varepsilon_i$), so that the energy associated with
the fluctuation $\delta{\bf u}$ is
\begin{eqnarray}
\Delta E(\delta{\bf u}) &=& \sum_{i=1}^{N_e} \left\{ \varepsilon_i
(w^\prime_\text{eff}[\rho_\text{in}, {\bf B}^{\bot}_\text{in} ])
- \varepsilon_i (w^\prime_\text{eff}[\rho_0, 0 ]) \right\}
\nonumber
\\
&& + {\cal O}_2(\delta n, \delta m,\delta{\bf B}^\bot ) ,
\end{eqnarray}
where ${\cal O}_p(x,y,\dots )$ is a quantity of order $p$ (and
higher) in $x,y,\dots$, and $\delta n \equiv n^\star - n_0$,
$\delta m \equiv m^\star - m_0$, $\delta {\bf B}^\bot \equiv {\bf
B}^{\bot\star} - {\bf B}^\bot_\text{in}$, respectively. We are
looking for an expansion of $\Delta E(\delta{\bf u})$ to second
order in $\delta{\bf u}$ of the form
\begin{equation}\label{eq_def_A}
\Delta E(\delta{\bf u}) = \sum_{\bf R,R^\prime} A_{\bf R
R^\prime}\delta {\bf u}_{\bf R}\cdot \delta{\bf u}_{\bf R^\prime}
+{\cal O}_4(\delta{\bf u}) ,
\end{equation}
with
%
$A_{\bf R R^\prime} \equiv  - J_{\bf R R^\prime} + \delta_{\bf
RR^\prime} \left( \sum_{\bf R^{\prime\prime}} J_{\bf R
R^{\prime\prime}} \right)$,
%
which {\em defines} the coupling parameters $J_{\bf RR^\prime}$.
The sum rule
\begin{equation}\label{eq_sum_rule}
\sum_{\bf R} A_{\bf RR^\prime} = \sum_{\bf R^\prime} A_{\bf
RR^\prime} = 0
\end{equation}
expresses the fact that the total energy remains invariant
(within the non-relativistic theory) upon a uniform rotation of
the magnetization. The definition (\ref{eq_def_A}) for $A_{\bf
RR^\prime}$ implies that it is related to the (static) transverse
susceptibility $\chi$ by: $2A_{\bf RR^\prime} = M_{\bf R} \left(
\chi^{-1} \right)_{\bf RR^\prime} M_{\bf R^\prime}$.

The {\em bare} MFT of Liechtenstein {\em et al.}
\cite{Liechtenstein1984} amounts to make the choice ${\bf
B}^\bot_\text{in} \equiv 0$ for the trial input constraining
fields. They then obtain
\begin{equation}
\Delta E(\delta{\bf u}) = \sum_{\bf R,R^\prime} \tilde{A}_{\bf R
R^\prime}\delta {\bf u}_{\bf R}\cdot \delta{\bf u}_{\bf R^\prime}
+{\cal O}_2(\delta n, \delta m, {\bf B}^{\bot\star}) ,
\end{equation}
with $\tilde{A}_{\bf R R^\prime} \equiv  - \tilde{J}_{\bf R
R^\prime} + \delta_{\bf RR^\prime} \left( \sum_{\bf
R^{\prime\prime}} \tilde{J}_{\bf R R^{\prime\prime}} \right)$, and

\begin{eqnarray}
\tilde{J}_{\bf R R^\prime} &\equiv& \frac{1}{\pi} \text{Im}
\int_{-\infty}^{\varepsilon_F}\!\!\! {\rm d}\varepsilon
\int_{\Omega_{\bf R}}\!\!\! {\rm d}{\bf r} \int_{\Omega_{\bf
R^\prime}}\!\!\! {\rm d}{\bf r^\prime} B_\text{xc}({\bf r})
G^\uparrow({\bf r},{\bf r^\prime})
\nonumber \\
&&\times B_\text{xc}({\bf r^\prime})G^\downarrow({\bf
r^\prime},{\bf r}) .
\end{eqnarray}
Since $\delta n$ and $\delta m$ are even with respect to
$\delta{\bf u}$, they are generally of second order in
$\delta{\bf u}$; however, the constrains ${\bf B}^{\bot\star}$
are odd with respect to $\delta{\bf u}$, and therefore generally
of first order in $\delta{\bf u}$, so that ${\cal O}_2(\delta n,
\delta m, {\bf B}^{\bot\star}) = {\cal O}_2 (\delta {\bf u})$,
although ${\cal O}_2(\delta n, \delta m) = {\cal O}_4 (\delta {\bf
u})$. As a consequence, the parameters $\tilde{J}_{\bf R
R^\prime}$ (hereafter called the {\em bare} exchange parameters)
are {\em not} equal to the {\em true} exchange parameters $J_{\bf
RR^\prime}$ and entail some systematic error. Note, however, that
this problem does not occur for the case considered by Oswald
{\em et al.} \cite{Oswald1985} since the constrains vanish in the
case they consider.

Clearly, to calculate correctly the {\em true} exchange parameters
$J_{\bf RR^\prime}$, we need to take the {\em exact} constrains
${\bf B}^{\bot\star}$ (which are still unknown yet) as trial
input in our estimate of the HF. If we do so, we obtain after
some algebra,
\begin{widetext}
\begin{equation}
\Delta E(\delta{\bf u}) = \sum_{\bf R,R^\prime}\left[
\tilde{A}_{\bf R R^\prime}\delta {\bf u}_{\bf R}\cdot \delta{\bf
u}_{\bf R^\prime} + (M_{\bf R}\delta_{\bf RR^\prime} -
\tilde{K}_{\bf RR^\prime}) {\bf B}^{\bot\star}_{\bf R} \cdot
\delta{\bf u}_{\bf R^\prime} - \frac{1}{2} \tilde{\chi}_{\bf
RR^\prime} {\bf B}^{\bot\star}_{\bf R} \cdot {\bf
B}^{\bot\star}_{\bf R^\prime} \right] +{\cal O}_4(\delta{\bf u}) ,
\end{equation}
with $M_{\bf R}\equiv \int_{\Omega_{\bf R}} {\rm d}{\bf r}\,
m({\bf r})$ is the magnetic moment of atom ${\bf R}$, and where
the {\em bare} transverse susceptibility $\tilde{\chi}_{\bf R
R^\prime}$ and the exchange-correlation response function
$\tilde{K}_{\bf R R^\prime}$ are given by
\begin{eqnarray}
\tilde{\chi}_{\bf R R^\prime} &\equiv& \frac{2}{\pi}
\int_{-\infty}^{\varepsilon_F}\!\!\! {\rm d}\varepsilon
\int_{\Omega_{\bf R}}\!\!\! {\rm d}{\bf r} \int_{\Omega_{\bf
R^\prime}}\!\!\! {\rm d}{\bf r^\prime}\ \text{Im} \left[
G^\uparrow({\bf r},{\bf r^\prime}) G^\downarrow({\bf
r^\prime},{\bf r})\right] , \\
\tilde{K}_{\bf R R^\prime} &\equiv&
\frac{1}{\pi} \int_{-\infty}^{\varepsilon_F}\!\!\! {\rm
d}\varepsilon \int_{\Omega_{\bf R}}\!\!\! {\rm d}{\bf r}
\int_{\Omega_{\bf R^\prime}}\!\!\! {\rm d}{\bf r^\prime} \
\text{Im}\left[ G^\uparrow({\bf r},{\bf r^\prime})
B_\text{xc}({\bf r^\prime})G^\downarrow({\bf r^\prime},{\bf r}) +
G^\downarrow({\bf r},{\bf r^\prime}) B_\text{xc}({\bf
r^\prime})G^\uparrow({\bf r^\prime},{\bf r})  \right].
\end{eqnarray}
\end{widetext}
The constrains are obtained by expressing the transverse moments
$M_{\bf R}\delta {\bf u}_{\bf R}$ as resulting from the
transverse exchange-correlation and constraining fields:
%
$
M_{\bf R}\delta {\bf u}_{\bf R} = \sum_{\bf R^\prime} \left(
\tilde{K}_{\bf R R^\prime} \delta{\bf u}_{\bf R^\prime} +
\tilde{\chi}_{\bf R R^\prime} {\bf B}^{\bot\star}_{\bf R^\prime}
\right) .
$
%
In order to keep the expressions compact we introduce matrix
notations: ${\sf A}$, ${\sf \tilde{A}}$, ${\sf \tilde{K}}$, ${\sf
\tilde{X}}$, ${\sf \tilde{M}}$ are the matrices whose $({\bf
RR^\prime})$-elements are, respectively, $A_{\bf RR^\prime}$,
$\tilde{A}_{\bf RR^\prime}$, $\tilde{K}_{\bf RR^\prime}$,
$\tilde{\chi}_{\bf RR^\prime}$, $M_{\bf R} \delta_{\bf
RR^\prime}$. Inserting the resulting expression of the constrains,
%
$
{\bf B}^{\bot\star}_{\bf R} = \sum_{\bf R^\prime} \left[ {\sf
X}^{-1} ({\sf M-\tilde{K}})\right]_{\bf RR^\prime} \delta {\bf
u}_{\bf R^\prime} ,
$
%
into the above expression of $\Delta E(\delta{\bf u})$, we
finally obtain the {\em exact} explicit expression of the {\em
renormalized} exchange parameters:
\begin{equation}\label{eq_renorm}
{\sf A}={\sf\tilde{A}}+\frac{1}{2} ({\sf M -\tilde{K}}^T )
{\sf\tilde{X}}^{-1} ({\sf M -\tilde{K}}) .
\end{equation}
One can easily prove the sum rule
%
$M_{\bf R} = \sum_{\bf R^\prime} \tilde{K}_{\bf RR^\prime}$,
%
which implies that the constrains ${\bf B}^{\bot\star}_{\bf R}$
vanish for a coherent rotation of all magnetic moment, and that
the sum rule (\ref{eq_sum_rule}) is indeed satisfied. The above
result, Eq.~(\ref{eq_renorm}), the ``{\em renormalized} magnetic
force theorem'', constitutes the main result of this Letter. Its
importance is that it corrects the systematic error introduced by
the ``{\em bare} magnetic force theorem'', without increasing
significantly the computational effort. It therefore provides a
general method to calculate {\em exactly} the exchange parameters
$J_{\bf RR^\prime}$ which is several orders of magnitude faster
than a fully self-consistent calculation.

The linearized equation of motion of the transverse fluctuations
is
\begin{equation}
M_{\bf R} \frac{{\rm d}\delta{\bf u}_{\bf R}}{{\rm d}t} = 2\,
\frac{\partial \Delta E}{\partial \delta{\bf u}_{\bf R}}\times
{\bf u}^0 = 4 \sum_{\bf R^\prime}A_{\bf RR^\prime} \delta{\bf
u}_{\bf R^\prime}\times {\bf u}^0 ,
\end{equation}
and the spin-wave energies are given, as usual, by the eigenvalues
of the symmetric matrix
%
$
\hbar {\sf \Omega} \equiv 4\, {\sf M}^{-1/2} {\sf A} {\sf
M}^{-1/2} .
$
%

Some better physical insight into the nature of the above
renormalization of the exchange parameters can be obtained if one
performs a simple, but yet quite reasonable, approximation. Let
us define
\begin{equation}
\Delta_{\bf R} \equiv \frac{2}{M_{\bf R}} \int_{\Omega_{\bf R}}
\!\!\! {\rm d}{\bf r}\, B_\text{xc}({\bf r}) m({\bf r}) =
\frac{4}{M_{\bf R}} \sum_{\bf R^\prime} \tilde{J}_{\bf RR^\prime}
,
\end{equation}
which can be seen as some average of the exchange splitting on
site ${\bf R}$. The second equality, in the above equation
expresses a sum rule related to the invariance with respect to a
global spin rotation. If the magnetization, within an atomic
cell, is sufficiently rigid, i.e., if intra-atomic fluctuations
of the spin-density have a large energy cost (as is usually the
case in transition metals), one has approximately:
\begin{equation}\label{eq_approx}
\tilde{K}_{\bf RR^\prime} \approx \frac{4\tilde{J}_{\bf
RR^\prime}}{\Delta_{\bf R}}
 , \text{ and }
\tilde{\chi}_{\bf RR^\prime} \approx \frac{8\tilde{J}_{\bf
RR^\prime}}{\Delta_{\bf R}\Delta_{\bf R^\prime}}  .
\end{equation}
Note that the above relations become exact if  do not perform the
discretization approximation ${\bf u(r) \to u_R}$ (see footnote
\cite{generalization_continuum}). One then obtains the following
simple and transparent renormalization for the exchange
interaction parameters,
\begin{equation}\label{eq_renorm_A}
{\sf A}= {\sf \tilde{A}} \left( 1- 4{\sf
M}^{-1}{\sf\Delta}^{-1}{\sf\tilde{A}} \right)^{-1},
\end{equation}
and for the spin-wave matrix
\begin{equation}
\hbar{\sf \Omega}= \hbar{\sf \tilde{\Omega}} \left( 1-
{\sf\Delta}^{-1} \hbar{\sf \tilde{\Omega}}  \right)^{-1} ,
\end{equation}
where $({\sf \Delta})_{\bf RR^\prime}\equiv \Delta_{\bf R}
\delta_{\bf RR^\prime}$ and $\hbar{\sf \tilde{\Omega}}\equiv 4\,
{\sf M}^{-1/2} {\sf \tilde{A}} {\sf M}^{-1/2}$ is the {\em bare}
(unrenormalized) spin-wave matrix. Noting that the bare
parameters $\tilde{A}_{\bf RR^\prime}$ can be expressed in terms
of the Stoner parameters $\left({\sf I}^{\text{xc}} \right)_{\bf
RR^\prime} \equiv \delta_{\bf RR^\prime} \Delta_{\bf R} / (2
M_{\bf R})$ as $2\tilde{\sf A} = {\sf M} {\sf I}^{\text{xc}}
\left( 1- \tilde{\sf X} {\sf I}^{\text{xc}} \right) {\sf M} $,
one easily shows that the renormalization (\ref{eq_renorm_A}) can
be re-expressed as
\begin{equation}
2{\sf A} \equiv {\sf M X^{-1} M} = {\sf M} \left( \tilde{\sf
X}^{-1} - {\sf I}^{\text{xc}} \right) {\sf M}.
\end{equation}
The above expression has the familiar form of the random-phase
approximation (RPA) result for the transverse susceptibility; it
is important to realize, however, that in the present context (as
shown by the above derivation) this result is {\em exact} within
the LDA (except for the discretization approximation ${\bf u(r)
\to u_R}$, which as already indicated, can be removed easily). The
present approach is therefore formally equivalent to approaches
based on calculations of the transverse susceptibility
\cite{Grotheer2001}, however without the need for self-consistent
total energy calculations.

For the particular case of periodic lattices, the above equations
are most conveniently solved in Fourier space. For systems with a
single atom per unit cell, and using the approximation
(\ref{eq_approx}), the renormalization of the exchange parameters
leads to the simple rescaling of the spin-wave spectrum
\begin{equation}
\hbar \omega({\bf q}) = \frac{\hbar \tilde{\omega}({\bf q})}{1-
\hbar \tilde{\omega}({\bf q}) / \Delta } .
\end{equation}
The above result clearly shows that the {\em bare} MFT yields
correct results only in the limit of long wavelength (in
particular, it yields the correct spin-wave stiffness $D$), or if
the spin-wave energies remain much smaller than the exchange
splitting $\Delta$. The above result is in agreement with the
estimate of the error entailed by the  {\em bare} MFT proposed by
Grotheer \cite{Grotheer2001}: $\hbar\tilde{\omega}\le
\hbar{\omega} \le
\hbar\tilde{\omega}/(1-\hbar\tilde{\omega}/\Delta)^{2}$.

The Curie temperature can be calculated by means of the
random-phase-approximation (RPA) Green's function method
\cite{Pajda2001}. For periodic systems with a single atom per unit
cell, one has
\begin{equation}
\frac{1}{k_B T_C^\text{RPA}} = \frac{6}{M}\, \frac{1}{N}\sum_{\bf
q} \frac{1}{\hbar \omega ({\bf q})} ,
\end{equation}
so that, by using approximation (\ref{eq_approx}), one therefore
obtains an extremely simple renormalization of the (RPA) Curie
temperature:
\begin{equation}\label{eq_Tc}
k_BT_C^\text{RPA} = k_B\tilde{T}_C^\text{RPA} \left( 1-6
\frac{k_B\tilde{T}_C^\text{RPA}}{M\Delta} \right)^{-1} ,
\end{equation}
where $\tilde{T}_C^\text{RPA}$ is the Curie temperature obtained
from the {\em bare} exchange parameters. As seen from
Table~\ref{tab_Tc}, the renormalization of the exchange parameters
considerably improves the agreement between theoretical and
experimental Curie temperatures of Fe and Ni.

The method discussed here provides a convenient and accurate
approach to study the exchange interactions, spin-wave spectra
and Curie temperature of complex systems such as disordered
alloys, ultrathin films, nanostructures, dilute magnetic
semiconductors (e.g., Ga$_{1-x}$Mn$_x$As), etc.

\begin{acknowledgments}
I wish to express my thanks to Helmut Eschrig for some very
stimulating discussions which motivated me to address this
problem. I am grateful to Josef Kudrnovsk\'y, Leonid Sandratskii,
and Valeri Stepanyuk for their helpful remarks on this work.
\end{acknowledgments}

%
\begin{table}[b]
\caption{\label{tab_Tc}Curie temperature calculated within the
RPA by using the {\em bare} ($\tilde{T}_C^\text{RPA}$) and {\em
renormalized} ($T_C^\text{RPA}$) exchange parameters, as compared
with the experimental value ($T_C^\text{exp}$). The {\em bare}
Curie temperature ($\tilde{T}_C^\text{RPA}$) is taken from
Ref.~\cite{Pajda2001}; the {\em renormalized} Curie temperature
($T_C^\text{RPA}$) is obtained from Eq.~(\ref{eq_Tc}).}
\begin{ruledtabular}
\begin{tabular}{lccc}
system & $\tilde{T}_C^\text{RPA}$ (K)  & $T_C^\text{RPA}$ (K)  &
$T_C^\text{exp}$ (K)  \\
\hline
Fe (bcc)   &   950   &   1057   &   1045 \\
Ni (fcc)   &  350    &   634   &   621 $-$ 631
\end{tabular}
\end{ruledtabular}
\end{table}

\vspace*{-1.0\baselineskip}


\end{document}